\documentclass[aps,prb,amsmath,amssymb,twocolumn,showpacs,10pt,superscriptaddress]{revtex4-2}
\usepackage[T2A]{fontenc}
\usepackage[utf8]{inputenc}
\usepackage{graphicx}
\usepackage{esint}
\usepackage[english]{babel}
\usepackage{xcolor}
\usepackage{hyperref}

\def\ext{\textrm{ext}}
\def\ed{\textrm{e}}
\def\g{\textrm{g}}
\def\gr{\textrm{gr}}
\def\q{{\tilde q}}

\begin{document}

\title{Quantum computation at the edge of a disordered Kitaev honeycomb lattice}

\author{Igor Timoshuk}
\affiliation{Condensed-matter physics Laboratory, HSE University, 101000, Moscow, Russia}
\author{Konstantin Tikhonov}
\affiliation{L. D. Landau Institute for Theoretical Physics, 142432 Chernogolovka, Russia}
\author{Yuriy Makhlin}
\affiliation{Condensed-matter physics Laboratory, HSE University, 101000, Moscow, Russia}
\affiliation{L. D. Landau Institute for Theoretical Physics, 142432 Chernogolovka, Russia}

\begin{abstract}
We analyze propagation of quantum information along chiral Majorana edge states in two-dimensional topological materials. The use of edge states may facilitate the braiding operation, an important ingredient in topological quantum computations. For the edge of the Kitaev honeycomb model in a topological phase, we discuss how the edge states can participate in quantum-information processing, and consider a two-qubit logic gate between distant external qubits coupled to the edge. Here we analyze the influence of disorder and noise on properties of the edge states and quantum-gate fidelity. We find that realistically weak disorder does not prevent one from implementation of a high-fidelity operation via the edge.
\end{abstract}

\maketitle
\section{Introduction}

Topological phases of matter may support boundary states with non-abelian statistics due to bulk-boundary correspondence. These states and their topological protection are useful in quantum-information processing~\cite{KitaevTopQC97,NayakReview}.
Recently, propagating Majorana edge modes in two-dimensional topological materials attracted attention~\cite{Zhang2010,Zhang2011,Zhang2015,Chen2017,Nagaosa2019}. Various methods to detect and characterize these modes were studied, including electrical probes of neutral Majorana zero modes by edge-state interferometry~\cite{AasenPRX20},
the use of time-domain interferometry to probe the edge modes and analysis of energy transport between external spins along the edge~\cite{KlockePRL21}, application of spin-polarized scanning tunneling microscopy to probe the charge-neutral edge states in Kitaev materials and other two-dimensional quantum magnets~\cite{Feldmeier}, or optical methods to probe chirality~\cite{Tero2022}.

Properties of chiral Majorana edge modes may be relevant for quantum-computing applications.
Topological quantum computations rely on braiding of non-abelian anyons as an elementary quantum logical gate, which is topologically protected. Since implementation of braiding of point-like anyons is an experimentally challenging task, it was suggested that the use of chiral one-dimensional edge modes may facilitate this step. It was demonstrated~\cite{Lian} that proper design of the edge interconnections may allow for braiding of Majorana fermionic excitations. This approach can be used to transfer quantum information along the edges and extended to realize quantum logical gates, in particular, in a Kitaev material~\cite{TM}. However, realization of materials, described by the Kitaev honeycomb or similar models is a difficult and actively investigated problem. Approaches include search for natural materials~\cite{Trebst} or design of artificial circuits, for instance, of quantum bits. In either approach the system may be subject to disorder and noise effects. This is especially relevant for qubit circuits, since artificial atoms (qubits) cannot be fabricated identical, and their couplings also vary between inter-qubit links. Understanding the effects of these imperfections on the physics of the edge states and their dynamics is important.

Here we focus on Majorana zero edge modes in the $B$-phase of the Kitaev honeycomb model~\cite{KITAEV20062} in a magnetic field. Possible realizations include Kitaev materials~\cite{Trebst} or artificial qubit networks with carefully tuned circuit parameters~\cite{Nori,Sameti,TM}. However, disorder and time-dependent noise may affect properties of the edge states and quality of quantum operations performed. These issues are analyzed in this article. In particular, we study stability of the flux-free energy sector under disorder and localization of edge states.
Further, we describe methods to transfer quantum information using the chiral edge states~\cite{TM} and study how various imperfections affect the quality of this operation and the corresponding quantum gates. Our results, on one hand, characterize the parameter ranges, suitable for needed quantum manipulations, and on the other hand may be used to probe various properties of the system.

\section{Exchange operations between external spins}
\label{sec:algo}

\begin{figure}[h]
\includegraphics[width=0.7\columnwidth]{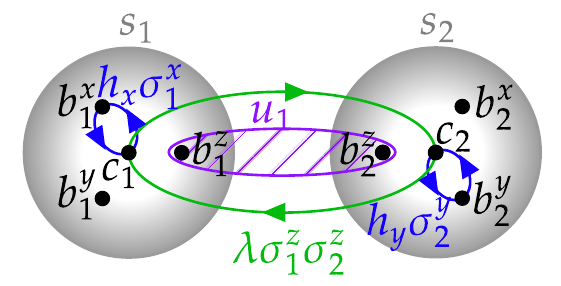}
\caption{\label{fig:Two-qubit-inter}
Outline of a Majorana-fermion exchange operations for two spins $s_1$ and $s_2$. Spin operators are represented by four Majoranas per spin. Exchange of fermions from different spins (green arrows) is achieved via the interspin coupling~\eqref{eq:zz_operation}, while the intra-spin Majorana exchange via the local magnetic fields~\eqref{eq:braiding_f} (blue).}
\end{figure}

The Kitaev honeycomb spin model~\cite{KITAEV20062} in external magnetic field $\mathbf{h}$ is defined via the Hamiltonian
\begin{equation}\label{eq:Ham}
\begin{gathered}
H=-J_x \sum_{x\textrm{-links}} \sigma_i^x \sigma_j^x - J_y \sum_{y\textrm{-links}} \sigma_i^y \sigma_j^y- \\
- J_z \sum_{z\textrm{-links}} \sigma_i^z \sigma_j^z-\mathbf{h} \sum_j\boldsymbol{\sigma}_j
\end{gathered}
\end{equation}
with summations over links in three different directions, $x$, $y$, $z$, on the honeycomb lattice. In this paper we assume $J_x=J_y=J_z=J$.
Following Ref.~\cite{KITAEV20062}, we use a fermionic representation for the spin-1/2 operators in terms of four Majoranas $c_j$, $b^{x,y,z}_j$ per site $j$ (see Fig.~\ref{fig:Two-qubit-inter}):
\begin{gather}
\sigma_{j}^{x}=ib_{j}^{x}c_{j},\qquad\sigma_{j}^{y}=ib_{j}^{y}c_{j},\qquad\sigma_{j}^{z}=ib_{j}^{z}c_{j},
\label{eq:majorana_transform}
\end{gather}
subject to a constraint of the physical subspace,
\begin{equation}
D_{j}=b_{j}^{x}b_{j}^{y}b_{j}^{z}c_{j}=1.
\label{eq:physical_subspace_restriction}
\end{equation}
The products $u_{jk}=i b_j^\alpha b_k^\alpha$, defined for every link $[ij]$ in direction $\alpha = x,y,z$, commute with the Hamiltonian and with each other, so that Eq.~(\ref{eq:Ham}) reduces to a quadratic Hamiltonian in each sector of fixed $u_{jk}=\pm1$, with the coupling terms $\frac{i}{2}Ju_{jk}c_jc_k$.

\begin{figure}[h]
\includegraphics[width=0.8\columnwidth]{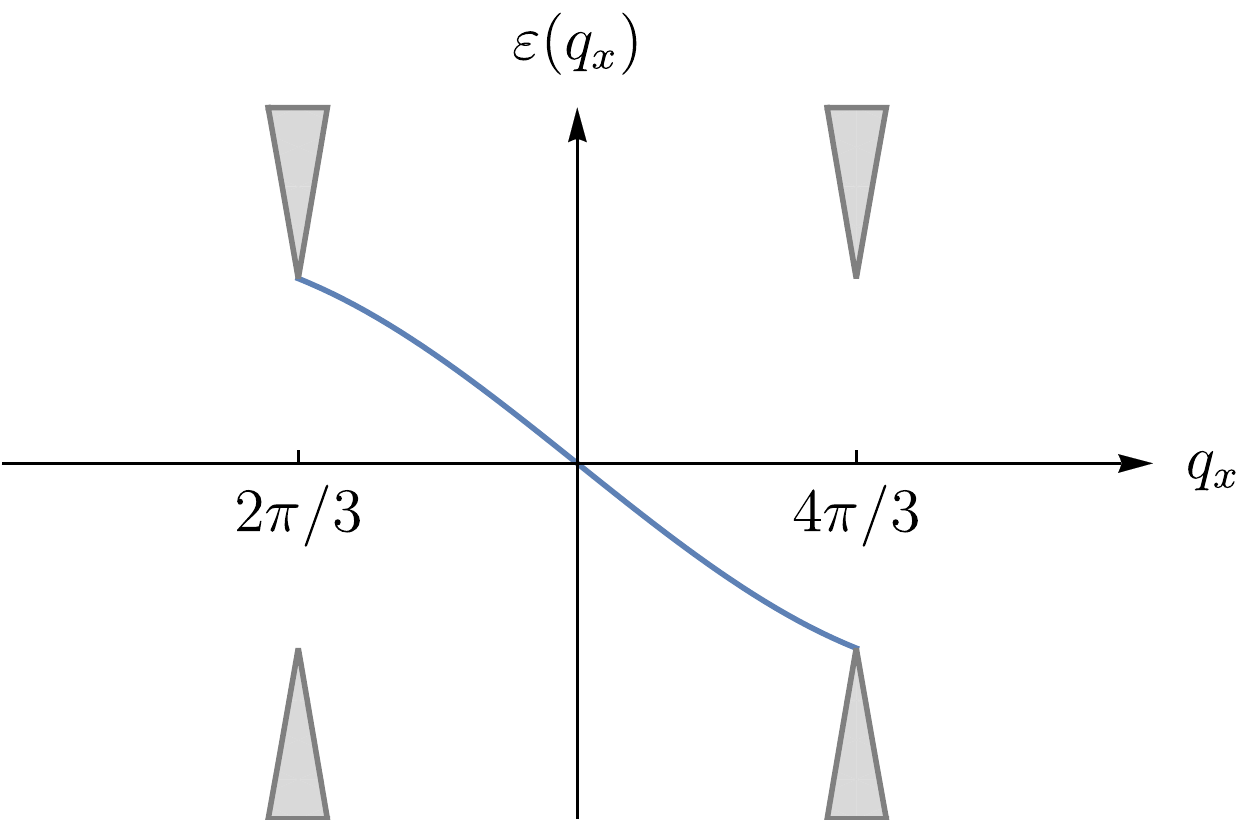}
\caption{\label{fig:zigzag_spect} Edge mode spectrum for a zigzag edge with zero $h_z$ component of the magnetic field at the boundary row of spins. Gray areas correspond to the continuous bulk spectrum.}
\end{figure}

Magnetic field breaks time-reversal symmetry and opens a gap in the bulk. In weak field, $h\ll J$, the effect of the field is described by the third-order contribution~\cite{KITAEV20062}:
\begin{equation}\label{eq:kappa}
V^{(3)} = -\kappa \sum_{jkl} \sigma^x_j\sigma^y_k\sigma^z_l \,,
\end{equation}
where summation is performed over triples $jkl$, in which one site is connected with the other two~\cite{KITAEV20062} and $\kappa\propto h_x h_y h_z/J^2$.
In the fermionic language, this term~(\ref{eq:kappa}) is also quadratic
and couples next-nearest neighbors, $\frac{i}{2}\kappa c_jc_l$.
Thus we obtain $c$-fermions on a honeycomb lattice with nearest- and next-nearest-neighbor couplings ($J$-terms and $\kappa$-terms)~\footnote{Cf.~Eq.~(48) in Ref.~\cite{KITAEV20062}.}

In the lowest-energy sector $u_{jk}=+1$ the system is translationally invariant~\cite{KITAEV20062,Lieb94}, and  in the momentum representation the Hamiltonian reads
\begin{align}\label{eq:HamSqFerm}
H &= \frac{1}{2} \sum_{\bf q} A({\bf q})_{\lambda\mu}
c_{-{\bf q}\lambda} c_{{\bf q}\mu} \,,\\
A({\bf q}) &=
\begin{pmatrix}\Delta({\bf q})&i f({\bf q})\\
-i f(-{\bf q})&-\Delta({\bf q})\end{pmatrix}
\,,\label{eq:A}\\
f({\bf q}) &= 2 J (e^{i{\bf qn}_1} + e^{i{\bf qn}_2}
+ 1)
\end{align}
where $\Delta ({\bf q}) = 4\kappa [\sin({\bf qn}_1)
+ \sin(-{\bf qn}_2) + \sin({\bf q}({\bf n}_2-{\bf n}_1))]$, ${\bf n}_{1,2} = (\pm1,\sqrt{3})/2$.
Here $\lambda$ and $\mu$ refer to the even or odd (black or white) sublattice.
Depending on the values of the coupling constants $J_{x,y,z}$ various phases can be realized~\cite{KITAEV20062}.  For $|J_x|<|J_y+J_z|$, $|J_y|<|J_x+J_z|$, $|J_z|<|J_x+J_y|$, the system is in the gapless $B$-phase, which will be of interest to us below. In this case, the gap in the spectrum in the absence of magnetic field closes at two opposite values of the momentum in the Brillouin zone, denoted as $\pm {\bf q}^*$.
The existence of these nodes is topologically protected by time-reversal symmetry (since under time reversal the structure of (\ref{eq:A}) persists).
Near the nodes $\pm{\bf q}^*$ the spectrum is parabolic:
\begin{equation}
\varepsilon({\bf q}) \approx
\pm\sqrt{3J^2\delta{\bf q}^2 + \Delta^2}
\end{equation}
with $\delta{\bf q}={\bf q}-{\bf q}^*$ (respectively, $\delta{\bf q}={\bf q}+{\bf q}^*$) and the gap $\Delta=6\sqrt{3}\kappa$.
Thus, the spectrum in the 2D-bulk is gapped, but due to the bulk-boundary correspondence a fermionic zero mode is expected at the boundary.

\begin{figure*}[ht!]
\includegraphics[width=0.9\textwidth]{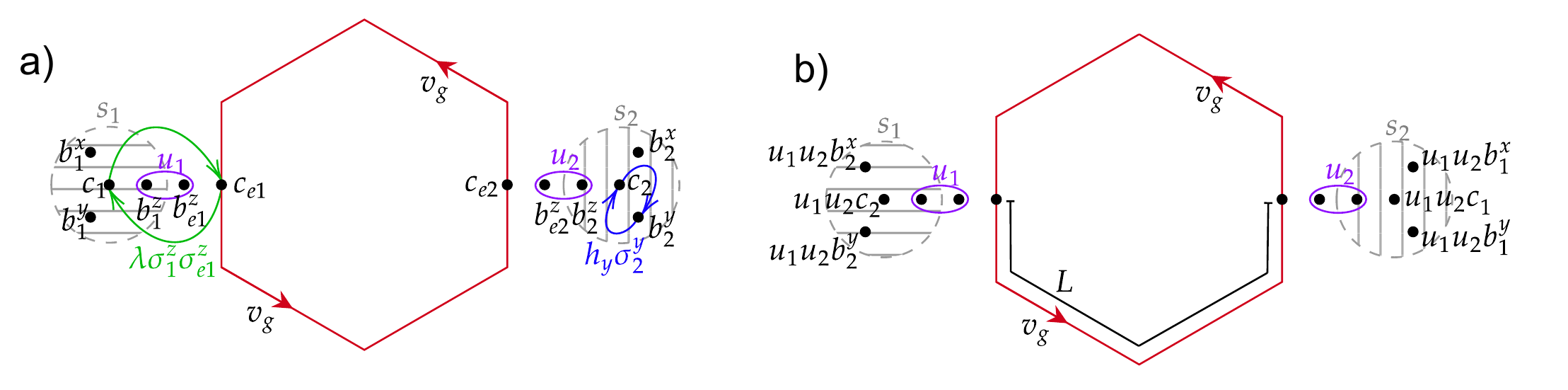}
\caption{\label{fig:swap_operation}
SWAP operation between two external qubits $s_1$ and $s_2$ at two different locations near the edge.
(a) Initial and (b) final~\eqref{eq:exchange_phys} locations of the constituent Majorana operators $c_i$, $b_i$.
Green and blue arrows indicate Majorana exchange operations between the qubits and within one qubit, respectively. These are performed via controlled spin couplings~\eqref{eq:zz_operation} and local fields~\eqref{eq:braiding_f}.}
\end{figure*}

Chiral Majorana edges can carry quantum states and can be coupled to external quantum systems. Manipulations which we will describe below can be applied for various edge types. However, properties of the edge states, their spectra and structure, may vary. It was argued in Ref.~\cite{TM} that the most suitable is the \emph{zigzag} edge of a Kitaev sample in uniform magnetic field, which vanishes at the edge, in the outermost row, $h_z=0$. This ensures conservation of the link operators $u_1$ and $u_2$ near both external spins.
While creation of this profile of magnetic field distribution may be challenging in `natural' Kitaev materials~\cite{Trebst}, it is much easier in artificial qubit lattices~(e.g., of superconducting qubits~\cite{Sameti,Nori}), where each coupling can be controlled individually, at least in principle~\cite{TM}.
In this approach, controllable $xx$, $yy$, and $zz$ interactions could be implemented~\cite{Nature99, Averin03, OliverCoupling18, Hutter06, RevModPhys.93.025005}.
In this configuration, the edge state of zero energy is localized near the edge~\cite{KITAEV20062}. 
Similar to Ref.~\cite{KITAEV20062}, a zero mode of the unperturbed Hamiltonian exists in the range $2\pi/3 < q_x <4\pi/3$ of longitudinal momenta (along the edge), see Fig.~\ref{fig:zigzag_spect}, and its wave function decays into the bulk with the decay factor $2\cos q_x/2$ per unit length. The spectrum of the edge mode obtained from the first-order perturbation theory,
\begin{equation}
\varepsilon = -12\kappa\sin q_x \,,
\end{equation}
is linear near zero energy with the velocity $v_\gr=-12\kappa$, cf.~Fig.~\ref{fig:zigzag_spect}.

We proceed to discuss how such systems may be used to initialize, transmit and readout quantum states. While our discussion is general, in our choice of needed manipulations we have in mind artificial qubit networks. In such systems local parameters may be controlled individually, at least in principle.

\subsection{Coupling external spins to edge states}

In this subsection we describe in the fermionic language how a two-qubit gate can be performed on two external spins coupled to the edge of a Kitaev honeycomb sample~\cite{TM}. In the process, quantum states are transferred between the spins using the chiral Majorana edge mode. The resulting operation is achieved in several steps, which include Majorana exchange operations, i.e., operations on a pair of Majorana operators, $\gamma_{1}$ and $\gamma_{2}$, with the result of a `$90^\circ$ rotation' in the $\gamma_1$-$\gamma_2$ plane:
\begin{equation}
\begin{gathered}U^{\dagger}\gamma_{1}U=\gamma_{2}\,,\\
U^{\dagger}\gamma_{2}U=-\gamma_{1}\,.
\end{gathered}
\label{eq:braiding_def}
\end{equation}
This operation can be effected by turning on the coupling Hamiltonian $H=ig\gamma_{1}\gamma_{2}$ for a  period $t=\pi/(4g)$.
In particular, local magnetic field
\begin{equation}
H_{h_{\alpha}}=h_{\alpha}\sigma_{j}^{\alpha}=ih_{\alpha}b_{j}^{\alpha}c_{j}
\label{eq:braiding_f}
\end{equation}
leads to an exchange \eqref{eq:braiding_def} of $c_j$ and $b_j^{\alpha}$ after a time span $t=\pi/\left(4 h_{\alpha}\right)$.

To exchange a pair of Majoranas from different spins, the spins should be coupled. For instance, the $zz$-coupling
\begin{equation}
H_{\lambda}=-\lambda\sigma_{1}^{z}\sigma_{2}^{z}=\lambda b_{1}^{z}c_{1}b_{2}^{z}c_{2}=\lambda\left(ib_{1}^{z}b_{2}^{z}\right)ic_{1}c_{2}
\label{eq:zz_operation}
\end{equation}
after a time span $\pi/\left(4\lambda\right)$ results in the exchange
\begin{equation}
\begin{gathered}U_{H_{\lambda}}^{\dagger} c_{1} U_{H_{\lambda}} = u_{12} c_{2} \,,\\
U_{H_{\lambda}}^{\dagger} c_{2}U_{H_{\lambda}} = -u_{12} c_{1} \,.
\end{gathered}
\label{eq:swap_u}
\end{equation}
of $c_1$ and $c_2$. Note the extra factor $u_{12} = ib_{1}^{z}b_{2}^{z} =\pm1$. Importantly, $u_{12}$ commutes with $H_{\lambda}$. 

\begin{figure*}[ht!]
\includegraphics[width=0.4\textwidth]{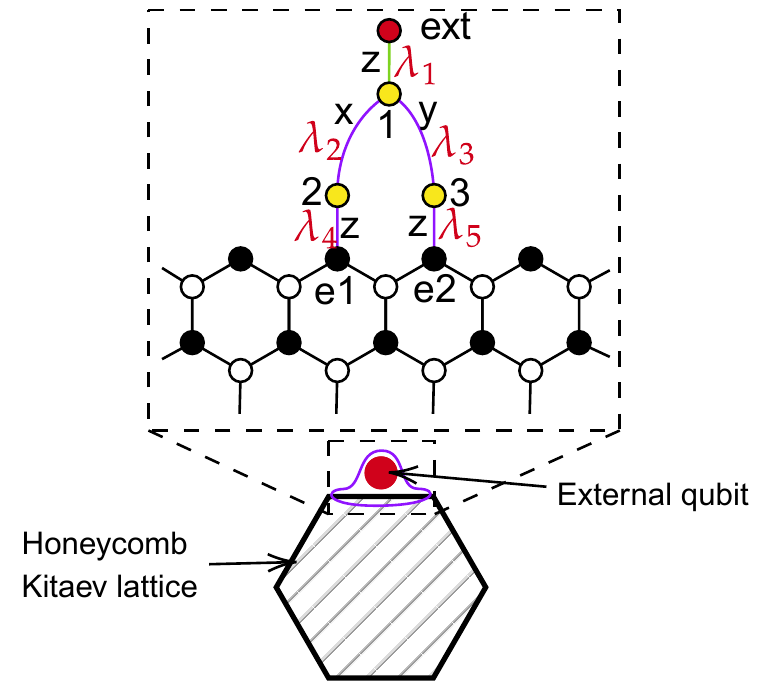}
\includegraphics[width=0.4\textwidth]{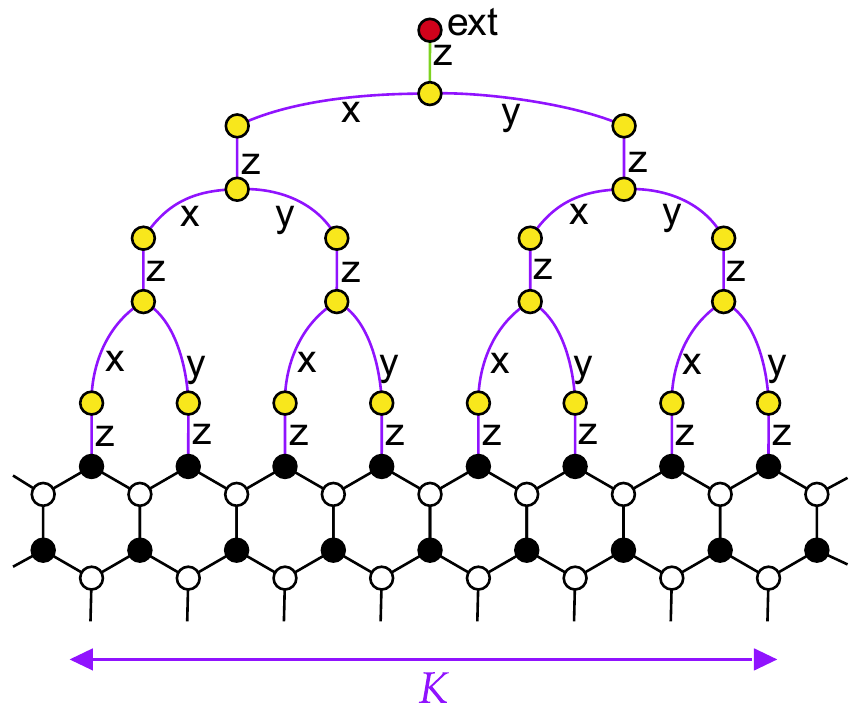}
\caption{\label{fig:Coupling_tree}Left: Interaction between an externally controlled qubit (red dot) and the edge of the Kitaev model lattice. Purple and green $\alpha$-edges correspond to $\sigma_{i}^{\alpha}\sigma_{j}^{\alpha}$ controllable interactions ($\alpha=x,y,z$).
Right: Extended construction, which allows to record and read out wider wave packets to/from the edge.}
\end{figure*}

Using these Majorana exchanges, one can construct an exchange of two spins (qubits) $s_{1}$ and $s_{2}$ using the edge states of the honeycomb model. 
We thus assume that the qubits are coupled to zigzag edges of the Kitaev honeycomb sample, see Fig.~\ref{fig:swap_operation} via the $zz$-coupling (this implies that the external spin is coupled via the $z$-component to the edge spin, which in turn is coupled to two other lattice spins via $x$- and $y$-links).

Applying external magnetic field, polarized along $x$ or $y$ directions, to the spins $s_{1}$ and $s_{2}$ we can exchange $b$ and $c$ Majoranas on each qubit~\eqref{eq:braiding_f} as indicated by the blue arrows in Fig.~\ref{fig:swap_operation}.
Turning on the interaction between an external spin and one of the edge spins, we achieve an exchange as in Eq.~\eqref{eq:swap_u}, which allows to record the state of the $c$-Majorana to the edge or read out the state from the edge to the $c$-Majorana.
Various write/read strategies can be used, and one needs to choose an optimal method depending on specific requirements and properties of the edge modes. For instance, one can keep the coupling $\lambda$ weak to achieve adiabaticity for the relevant bulk states ($\lambda\ll J$ for the range of states discussed below), at the same keeping it strong enough, so that the process is instantaneous for the low-energy edge state ($\lambda\gg v_\gr$). This would create a narrow wave packet. Alternatively, one can couple strongly, $\lambda\gg J$, ensuring instantaneous coupling for all states. In this case, one aims at writing a wave packet which does not overlap with the bulk states, but only with the edge modes. Furthermore, the spectrum of the edge modes is linear only at very low energies, thus in order to prevent strong deformation  of the travelling wave packets due to nonlinearities, one may aim at a sufficiently wide wave packet. One can achieve this by keeping a weak coupling on for a sufficiently long period~\cite{TM}, so that the wave packet moves during the writing (respectively, readout) operation. This allows one to achieve high fidelities~\cite{TM}, but prevents one from controlling the shape of the wave packet.
For completeness, in Section~\ref{sec:exch_details} we describe another approach, which enables creation of a wide wave packet with full control of its shape.
Note, however, that here we focus on propagation of quantum states along the edge and analyze fidelity of this process. From this viewpoint, a specific choice of the recording/readout procedure is not crucial here.

We outline first the full exchange operation, leaving some details to Section~\ref{sec:exch_details}: First, record the $c_1$-fermion of the 1st qubit to the edge, then wait until it propagates to the location of the 2nd  qubit, and perform a similar exchange (readout) operation at the location $s_2$. Further, similarly the fermions $b^x_1$ and $b^y_1$ can be transferred to spin 2.
As a result of these manipulations, operators $\sigma^\alpha_1 = i b_1^{\alpha} c_1$ and $\sigma^\alpha_2$ are exchanged for each $\alpha=x,y,z$ (see Fig.~\ref{fig:swap_operation}b):
\begin{align}
\sigma_{1}^{x} & \longleftrightarrow\sigma_{2}^{x},\nonumber \\
\sigma_{1}^{y} & \longleftrightarrow\sigma_{2}^{y},\label{eq:exchange_phys}\\
D_{1}\sigma_{1}^{z} & \longleftrightarrow D_{2}\sigma_{2}^{z},\nonumber 
\end{align}
where $D_{j}=b_{j}^{x}b_{j}^{y}b_{j}^{z}c_{j}$. In the physical subspace, $D_{1}=D_{2}=1$ (cf.~Ref.~\cite{KitaevTopQC97} and Section~\ref{sec:algo}), and hence exchange of the external qubits' states is achieved.

\subsection{Exchanging the external spin states via edge states}
\label{sec:exch_details}

Let us demonstrate how a state of an external qubit can be written to a wide wave packet at the edge. While point-like coupling to a single edge spin can produce a narrow packet, one may need a wider packet, for instance, to ensure that only long-wavelength edge modes are involved, and the relevant, low-energy part of the spectrum is linear, so that the wave packet is transmitted without distortion. To show that this is possible in principle and an arbitrary-shape wave packet may be created, we consider the setup in Fig.~\ref{fig:Coupling_tree}. With this choice of couplings, shown in the figure, the external yellow spins may be viewed as a part of the Kitaev lattice. In particular, the operators $u$ at the external purple links are conserved during operation. We have verified numerically that the ground state of such an augmented honeycomb lattice does not carry vortices unless the auxilary tree is large enough. Thus, we can work in the gauge, where the link operators are $u_{ij}=1$ at the external purple $\lambda_i$ links.

Let us first show that one can record the initial arbitrary state of the external (red) qubit onto the edge wave packet, which occupies two edge spins $e1$, $e2$, see Fig.~\ref{fig:Coupling_tree}, left. 
The initial state of an external (red) qubit is arbitrary, hence at the green link the value of $\hat{u}_{g}$ is arbitrary, similar to $u_{1}$ and $u_{2}$ in the discussion above. Initially, we keep the external $\lambda_i$ couplings turned off and prepare the red qubit in some initial state, which is to be transferred via the edge modes.

We now describe the operation in more detail. We assume that initially couplings at the blue and green edges are switched off.
The transfer a fermion operator $c_1$ from the red qubit to a wave packet at the edge is achieved in three steps by gradually transferring it row by row. First, we apply a pulse of the green $\lambda_1$ coupling for a period $t_1=\frac{\pi}{4\lambda_{1}}$
to exchange $c_\ext$ with $c_1$ at site 1:
\begin{equation}
c_\ext \mapsto u_\g c_1 \,.
\end{equation}
At the next step, the couplings $\lambda_2$ and $\lambda_3$ are switched on:
\begin{equation}
H=\lambda_{2}\sigma_{1}^{y}\sigma_{2}^{y}+\lambda_{3}\sigma_{1}^{x}\sigma_{3}^{x} \,.
\end{equation}
In the fermionic language and in the absence of vortices, $u_{ij}=1$, this Hamiltonian reads: 
\begin{equation}
H=i\lambda_{2}c_{1}c_{2}+i\lambda_{3}c_{1}c_{3}
= i\sqrt{\lambda_{2}^{2}+\lambda_{3}^{2}}c_{1}c_{23}
\,,
\end{equation}
where a new Majorana fermion is defined by $c_{23}=\frac{1}{\sqrt{\lambda_{2}^{2}+\lambda_{3}^{2}}}\left(\lambda_{2}c_{2}+\lambda_{3}c_{3}\right)$. Similar as above, such interaction, if switched on for $t_2=\frac{\pi}{4\sqrt{\lambda_{2}^{2}+\lambda_{3}^{2}}}$, exchanges the states of $c_1$ and $c_{23}$. As a result, $c_\ext$ of the external qubit propagates to the next row of spins: 
\begin{equation}
c_\ext \mapsto \frac{u_\g}{\sqrt{\lambda_2^2+\lambda_3^2}}
(\lambda_2c_2 +\lambda_3c_3) \,.
\end{equation}

Proceeding similarly in the following row, we switch on the interactions $\lambda_4$ and $\lambda_5$ (one may choose $\lambda_4=\lambda_5$) of the qubits 2 and 3 with the edge qubits for $t_3=\frac{\pi}{4\lambda_{4}}$, one more exchange is performed and finally we find that the initial $c_\ext$ is mapped to
\begin{equation}
c_\ext \mapsto
\frac{u_\g}{\sqrt{\lambda_{2}^{2}+\lambda_{3}^{2}}}
(\lambda_2 c_{\ed1} + \lambda_3 c_{\ed2}) \,.
\label{eq:edge_pac}
\end{equation}
Note that the shape of the resulting wave packet \eqref{eq:edge_pac} can be controlled via the couplings $\lambda_2$, $\lambda_3$.

After this recording procedure, this wave packet \eqref{eq:edge_pac} propagates along the edge and arrive at the location of the other (`primed') external qubit. In other words, after some travel time $t_\textrm{tr}$ we find that $c_{\ed i}\mapsto c'_{\ed i}$ near this primed qubit.
There one can swap the state of the wave packet onto this other external (red) qubit by following the same row-by-row procedure in reverse order.
Thus, the steps performed effect the sequence of operations:
\begin{eqnarray}
c_\ext &\mapsto& \frac{u_\g}{\sqrt{\lambda_2^2+\lambda_3^2}}
(\lambda_2c_{\ed1} +\lambda_3c_{\ed2}) \nonumber\\
&\mapsto&\frac{u_\g}{\sqrt{\lambda_2^2+\lambda_3^2}}
(\lambda_2c'_{\ed1} +\lambda_3c'_{\ed2})
\mapsto c'_\ext 
\,.
\end{eqnarray}

Thus, finally $c_\ext$ of the first external qubit is swapped onto $c'_\ext$ of the second external qubit via the wave packet at the edge as a mediator.
 
In reality, however, due to various imperfections (such as deformation of the wave packet due to nonlinearity of the edge-mode spectrum, influence of disorder or noise), the propagation is not perfect. Then the $c_{\ed i}\mapsto c'_{\ed i}$ mapping for $i=1,2$, implied by perfect propagation, is replaced by
$c_{\ed i} = P c'_{\ed i} + \sqrt{1-P^2} d_{\ed i}$, where the Majorana operator $d_{\ed i}$ is a normalized  ($d_{\ed i}^2=1$) superposition of Majoranas at other edge sites (so that $c'_{\ed i}$ and $d_{\ed i}$ anticommute). Thus, the initial $c$-Majorana at the first external qubit is mapped to that for the second external qubit with errors:
\begin{equation}\label{eq:overlap_d}
c_\ext \mapsto P u_\g u'_\g c'_\ext
+ \sqrt{1-P^2} d'_\ed \,,
\end{equation}
$d'_\ed$ being some combination of edge Majoranas around the location of the second qubit.

As we will see in Section~\ref{sec:fidelity}, the quantity $P$ determines the fidelity of the exchange operation. From the description above it is clear that the  value of $P(t)$ in a given configuration depends on the travel time along the edge, and the optimal value of $P(t)$ is attained at certain $t^*$, so that one should pick $t_{tr}$ near $t^*$. The suppression of $P(t)$ from its maximal value depends on the ratio $L\delta t/wt^*$ between the time mismatch $\delta t = t_{tr} - t^*$ and the width of the wave packet $w$, where $L$ is the distance between the external qubits along the edge, see Fig.~\ref{fig:swap_operation}.

The recording (and reading-out) algorithm, described above, can be generalized to enable creation of edge wave packets of a larger span and with arbitrary spatial profile. This requires larger tree-like structures, see Fig.~\ref{fig:Coupling_tree}, right. 

These structures allow one to create superpositions over $K$ edge sites
\begin{equation}\label{eq:lin_comb}
c_\ext \mapsto u_\g\sum_{i=1}^{K}\alpha_{i}c_{\ed i},
\qquad
\alpha_{i}\in\mathbb{R},\qquad\sum_{i=1}^{K}\alpha_{i}^{2}=1
\end{equation}
with the weights $\alpha_i$ determined by the purple couplings in the tree.

\begin{figure}[h!]
\includegraphics[width=0.8\columnwidth]{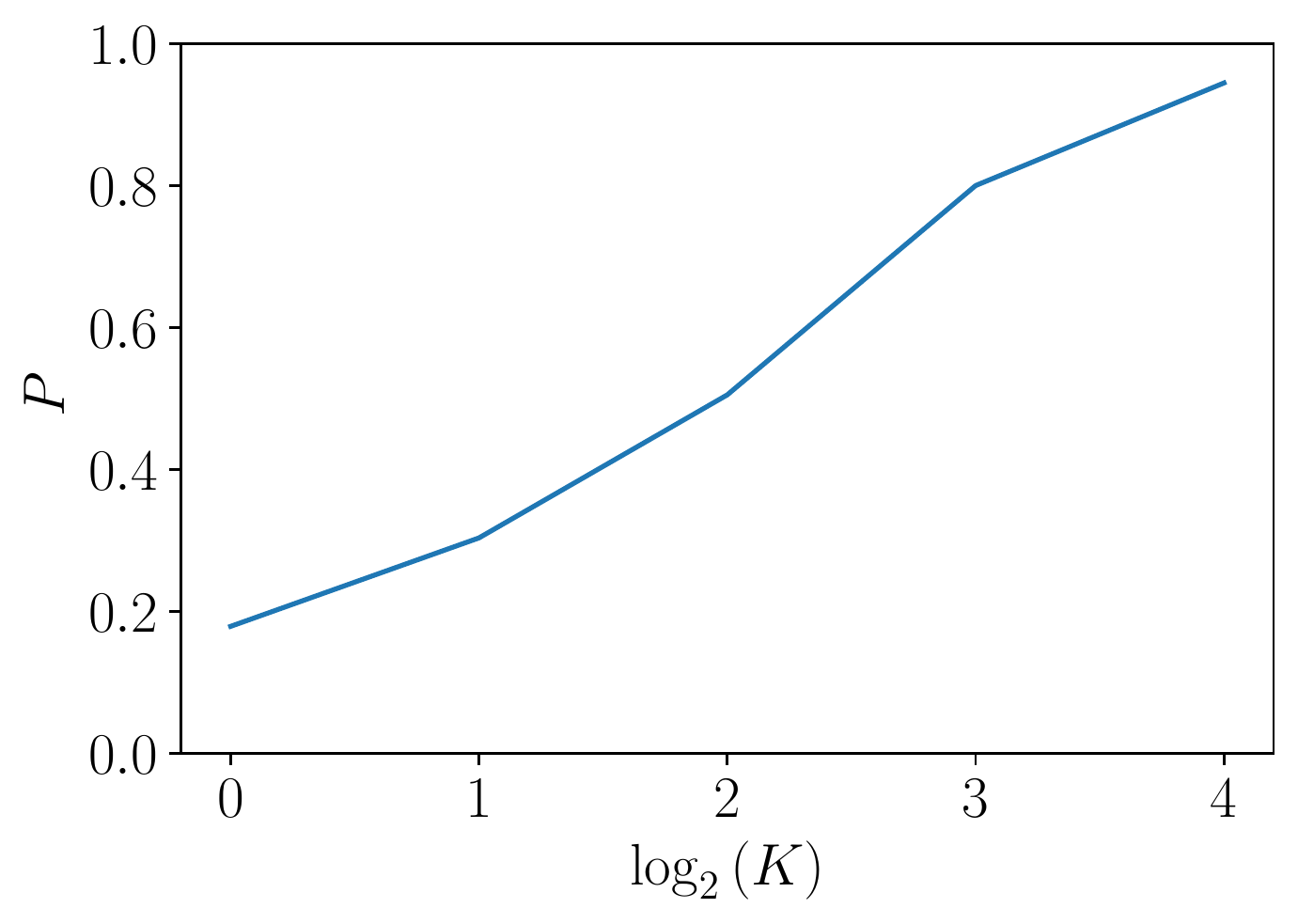}
\caption{\label{fig:4_data}
Overlap with the perfectly transmitted wave packet vs. width $K$ (see  Fig.~\ref{fig:Coupling_tree}).
Its suppression is caused by non-linearity of the edge-mode dispersion and is weaker for a wider wavepacket.
}
\end{figure}

For deeper tree-like structures, wider and hence less dispersing wave packets are created, which improves the overlap $P$, see numerical results in Fig.~\ref{fig:4_data}. In this figure, we illustrate suppression of the overlap caused by nonlinearity in the spectrum, $\varepsilon_0(q)-v_\gr^0q$, and imperfections in the write and read procedures from the coupling to the bulk states. Complexity of this construction grows with the size $K$ of the desired wave packet, however, due to parallelization, the required time grows only logarithmically.
We observed that the size of the tree is limited by its instability to vortex formation, and at fixed coupling strength a large enough tree hosts a vortex below its largest arc in the ground state.

Let us also remark that while the qubit-edge coupling described above allows for arbitrary-profile wave packets, applications to quantum-state transfer may also rely on other procedures, for instance, longer coupling via a point-like contact~\cite{TM}, which may be more suitable experimentally depending on qubit realization.

\subsection{Fidelity}
\label{sec:fidelity}

From the expression \eqref{eq:overlap_d} for the mapping of the $c$-fermion and similar expressions for transfer of $b^x$ and $b^y$, we find that $\sigma^\alpha_\ext = i b^\alpha_\ext c_\ext$ is transformed to
\begin{eqnarray} \label{eq:sstransfer}
\sigma^\alpha_\ext &\mapsto&
P^2 (\sigma'_\ext)^\alpha
+ P\sqrt{1-P^2} u_\g u'_\g (c' d_{b^\alpha}
\nonumber\\
&&+ b^{\prime\alpha} d_c) + (1-P^2) d_c d_{b^\alpha}
\,.
\end{eqnarray}
We note that the factors of $u_\g$ and $u'_\g$ cancel out from the first term. Thus in the limit of perfect edge Majorana transmission, $P=1$, we indeed achieve a perfect spin swap.
Further, from this expression one can evaluate the fidelity of the constructed spin-swap operation. If we are interested in the quality of transfer of the first qubit's state, one can consider its arbitrary initial mixed state $\rho$ and evaluate the final state of the second external qubit.

From Eq.~\eqref{eq:sstransfer} we find that
$\rho\mapsto \frac12 + P^2(\rho-\frac12)$.
We may use the Uhlmann-Josza fidelity~\cite{Uhlmann76,Josza94,Liang2019} for mixed states or the $A$-fidelity~\cite{Afidelity,Liang2019}, $F(\rho,\rho') = (\mathop{\textrm{Tr}}\sqrt{\rho\rho'})^2$, since $\rho$ and $\rho'$ commute. After averaging over possible initial states~\cite{Pedersen2007}, the resulting value of the fidelity of the constructed spin-swap operation is
\begin{gather}\label{eq:fidelity}
F = \frac{1+P^{2}}{2}.
\end{gather}
Below we characterize fidelity of operations via $P^2$.

\section{Fidelity at static disorder}
\label{sec:staticdisorder}

In this section we evaluate the contribution to decoherence, caused by $\delta$-correlated bond disorder. Such systems with periodic boundary conditions and various coupling strength distributions were discussed in \cite{PhysRevB.92.014403, PhysRevB.102.054437, KAO2021168506}. We assume that each nearest- and next-nearest-neighbor coupling in the lattice is sampled from a Gaussian distribution: with mean $J$ and variance $\langle\delta J_{ij}^2\rangle=\sigma^2 J^2$
and mean $\kappa$ and variance $\langle\delta \kappa_{ij}^2\rangle=\sigma^2 \kappa^2$, respectively.

\begin{figure} 
\includegraphics[width=0.8\columnwidth]{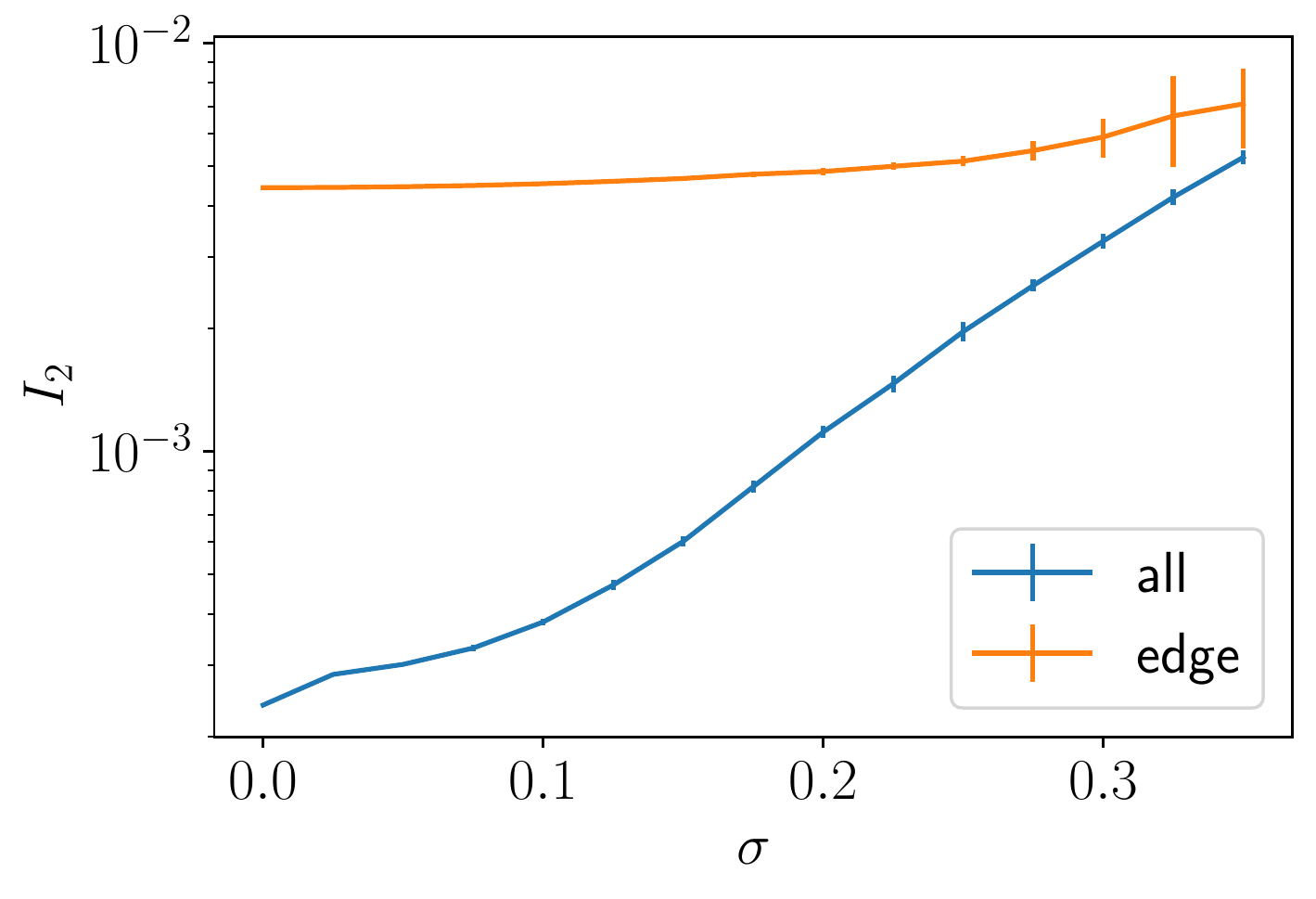}
\caption{\label{fig:IPR_log}Mean IPR of the edge and bulk states as a function of disorder, quantified via
the standard deviation of the coupling strength parameters.}
\end{figure}

Disorder may modify properties of the system in various aspects. First, we have to verify that moderate level of static disorder leaves the system remains in the vortex-free sector. As numerical simulations show, at least for $\sigma\lesssim 0.35$, this is indeed the case. Further, disorder may localize the Majorana edge modes. Due to their chiral nature \cite{KITAEV20062, TM}, they are expected to be robust towards localization at sufficiently weak disorder. Stability to localization can be verified by inspecting the inverse participation ratio (IPR)
\begin{equation}
\label{ipr_def}
I_2 = \left\langle\sum_k \sum_{i=1}^N \left|\psi^k_i\right|^4 \right\rangle \,,
\end{equation}
where $N$ is the number of nodes in the sample, $\psi^k_i$ is the amplitude of the $k$-th state at the $i$-th node. Averaging in Eq.~\eqref{ipr_def} is performed over eigenstates in a certain energy range and over disorder realizations. In order to verify the limits to the edge state stability, we numerically compute the mean IPR for all $N$ eigenstates of the disordered Hamiltonian and compare it with the mean IPR for the edge states. As the spectrum in Fig.~\ref{fig:zigzag_spect} suggests, the edge states lie in the gap of the bulk spectrum and can be easily identified by the corresponding eigenvalue. For a system of $N=9600$ nodes and perimeter of $L=240$ sites, we show the values of IPR in Fig.~\ref{fig:IPR_log}. According to these results, the edge states remain fully extended up to disorder levels $\sigma\lesssim 0.2$ (indeed, $I_2=1/L$ corresponds to a plane-wave-like eigenstate propagating along the boundary).

In our further analysis we assume that disorder is sufficiently weak so that the ground state is vortex-free edge modes are delocalized. Nevertheless, for the algorithm in Section~\ref{sec:algo}, it is crucial to read out the edge state at the optimal time, when the  wave packet arrives at the read-out position: possible mismatch suppresses fidelity of the operation. In this section, we evaluate the effect of static disorder on propagation of the edge state.

For simplicity, we assume that the write and read operations are performed on the same set of nodes, at the same position: in other words, the wave packet makes a full circle around the sample between these operations. The major effect of disorder is modification of the edge-mode velocity, which depends on a particular realization. Here we evaluate corrections to the velocity and its sample-to-sample fluctuations to characterize the effect of disorder on the fidelity$F$, see Eq.~\eqref{eq:fidelity}.

\begin{figure*}[ht!]
\includegraphics[width=0.45\textwidth]{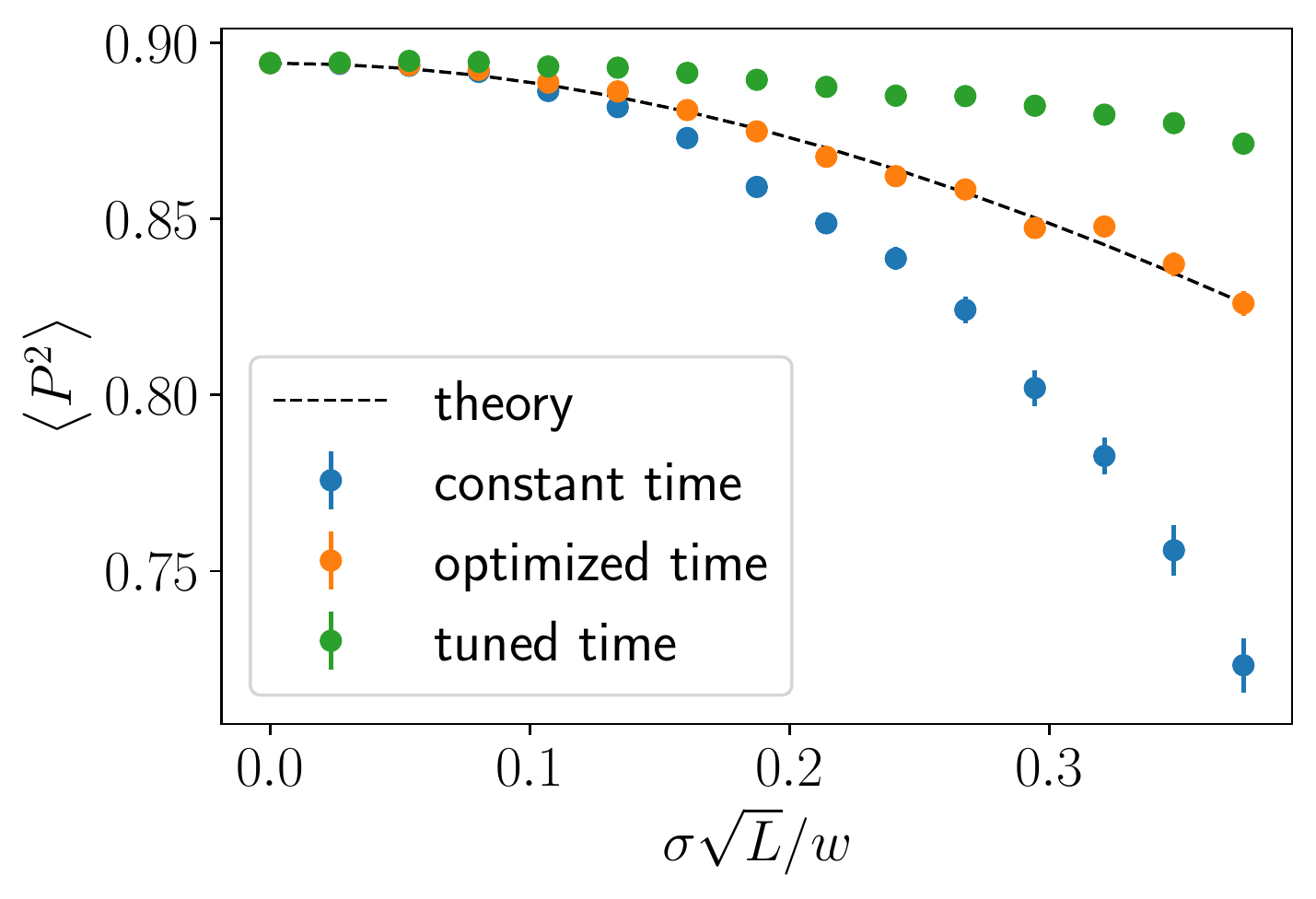}
\includegraphics[width=0.45\textwidth]{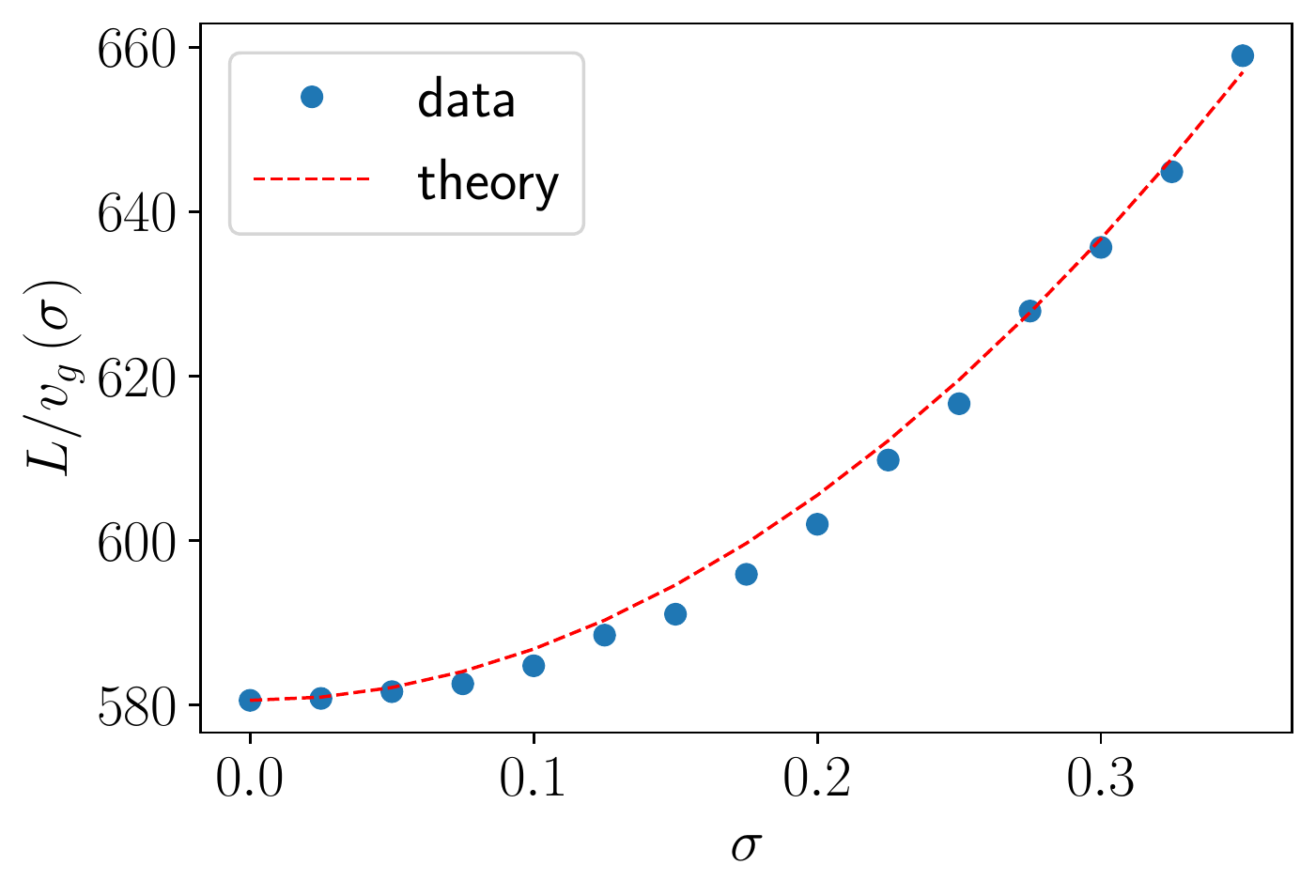}
\caption{\label{fig:Overlap_vs_disorder_const}
Left: The overlap between the initial and final wave packets vs. disorder. The overlap was evaluated numerically for three different readout methods: readout at the optimal travel time in the clean system, $t_{tr} = L/v_\gr(0)$ (\emph{constant});
readout at the time point, \emph{optimized} for the disorder-averaged velocity, $L/v_\gr(\sigma)$, plotted together with the expected value~\eqref{eq:overlap} (dashed curve);
and readout at time $t^*$ maximizing $P^2$ individually for each specific realization of disorder (\emph{tuned}).
At each point, the mean value and standard deviation are calculated for an ensemble of 300 random hexagon samples with 2400 nodes.
Right: Optimal overlap time for a hexagonal sample with 2400 nodes. The observed slowdown follows Eq.~\eqref{eq:delta_vg}.}
\end{figure*}

If disorder is not too strong, its main effect can be captured perturbatively. Every initial state can be represented as a linear combination of the eigenstates in the disordered system. First- and second-order corrections to the eigenenergies from disorder modify the dynamic phases of the eigenstates. Taking periodicity into account, the final overlap after one turn can be found as 
\begin{equation}\label{eq:overlapFourier}
P(t) = \sum_q A_q^2 e^{iqL - it(\varepsilon_0(q) + \delta\varepsilon(q) )}
\,,
\end{equation}
where $\varepsilon_0(q)$ is the spectrum in a clean system, $\delta\varepsilon(q) = \varepsilon_1(q) + \varepsilon_2(q) +\dots$ is the perturbative correction, and the shape of the wave packet is defined via its Fourier transform $A_q$.
The first-order correction $\varepsilon_1$ is given by:
\begin{equation}
\varepsilon_1(q_x) = V_{q_xq_x}
\end{equation}
via the diagonal matrix element $V_{q_xq_x}$ of the disorder-induced part $V$ of the Hamiltonian for the edge state with wavevector $q_x$. At $q_x=\pi+\q_x$ and small $\q_x$, the state is localized on the first row of nodes~\cite{KITAEV20062}, where the amplitude of the normalized state is $1/\sqrt{L}$. We find for the first-order disorder-induced energy correction:
\begin{equation}\label{eps_1_mom}
\langle \varepsilon_1(q_x)\rangle =0,\quad
\langle \varepsilon_1^2(q_x)\rangle = \frac{176\kappa^2}{L} \sigma^2 \q_x^2\,.
\end{equation}
In the second order of perturbation theory, the correction is non-zero and is dominated by transitions to the bulk states:

\begin{gather}
\left\langle \varepsilon_2(q_x)\right\rangle \approx 
-\varepsilon_0(q_x) \sigma^2 \cdot\frac{1}{N}
\sum_{\boldsymbol{q}'\in\textrm{BZ}} \frac{4J^2}{|E_{\boldsymbol{q}'}|^2}.
\end{gather}
Evaluating the sum, we find a correction to the average velocity of the edge mode (the numerical coefficient is quoted for $\kappa/J=0.027$):
\begin{equation}\label{eq:delta_vg}
\delta v_\gr =-1.07 v_\gr \sigma^2,
\end{equation}
which indicates that the wavepacket is slowed down, on average.


Let us now evaluate the overlap of the wavepackets after a full rotation.
As we discussed in Section~\ref{sec:exch_details}, the overlap is suppressed to a certain value $P_0$ already in a clean system due to nonlinearity of the spectrum and imperfections in coupling to the edge, see Fig.~\ref{fig:4_data}. Here we evaluate additional suppression due to static disorder. Disorder modifies the velocity at the edge and hence the final position of the wave packet at a certain time $t$. This random shift is determined, on average, by Eq.~\eqref{eq:delta_vg}, and fluctuates according to Eq.~\eqref{eps_1_mom}. One may try to optimize the overlap by adjusting the readout time $t$.

The average overlap squared, $P^2$, at the \emph{average} optimal readout time $t=L/(v_\gr^0+\delta v_\gr)$ becomes
\begin{equation}\label{eq:P2opt}
P^2(t) = P_0^2 \sum_{q,q'} A_q^2 A_{q'}^2
e^{-it (\delta \varepsilon(q) - \delta \varepsilon(q'))}
\,.
\end{equation}
For a Gaussian initial wave packet of width $w$, averaging Eq.~\eqref{eq:P2opt} over disorder, we find
\begin{equation}\label{eq:overlap}
\langle P^2\rangle = P_{0}^{2}
\left(1+\frac{11\sigma^2L}{9w^2}\right)^{-1/2} \,.
\end{equation}

Note that the disorder strength $\sigma$ should be compared to the size-dependent emergent scale $\sim w/\sqrt{L}$. Indeed, scattering at each random link shifts the travelling wave packet by $\sim\pm\sigma$, accumulating to $\sigma\sqrt{L}$ due to random signs.

We numerically evaluated the overlap $P^2$ between the original and final wave packets using three different read-out approaches, see Fig.~\ref{fig:Overlap_vs_disorder_const}. One method uses readout at the bare arrival time $L/v_\gr^0$ of the clean system. In another approach, optimization for the disorder-averaged delay of the wave packet by reading out at time $t=L/(v_\gr^0+\delta v_\gr)$ improves the fidelity. This optimized numerical fidelity is compared to Eq.~\eqref{eq:overlap} (dashed line). Finally, one can try to tune to the optimal readout time individually for each disorder configuration.

\section{Fidelity under non-stationary noise}

Consider now time-dependent noise in the system. We first analyze the effect of uniform non-stationary noise, assuming that each coupling in the system follows the same noisy pattern: $\delta J_{ij}(t)/J_0 = \delta \kappa(t)/\kappa_0 = \xi(t)$. In this case, each energy eigenstate evolves as follows:
\begin{equation}
\psi_\varepsilon(t) = \exp\left(-i \varepsilon \intop_0^t dt' (1 + \xi(t')) \right) \psi_\varepsilon(0) \,. 
\end{equation}
In the limit of low-frequency noise, all the couplings in the Kitaev model fluctuate collectively between experimental runs. Physically, such a situation may be realized if all the couplings are controlled by a uniform fluctuating field. 
The same effect occurs if the couplings are robust, but the waiting time $t_{\textrm{tr}}$ fluctuates. Thus such simple structure of fluctuations may nevertheless induce nontrivial consequences for the fidelity.

The overlap between the clean and noisy wave functions depends on the quantity $\chi(t) = \intop_{0}^{t} dt' \xi(t')$
and its statistical properties. Assuming Gaussian noise $\xi(t)$, we observe that $\chi(t)$ is also Gaussian distributed.
Its variance grows with time and can be found explicitly,
\begin{gather}
\langle \chi^2(t)\rangle =
\int \frac{d\omega}{2\pi} \langle\xi^2_\omega \rangle
\left(\frac{\sin(\omega t/2)}{\omega/2}\right)^2
\,,
\end{gather}
in terms of the spectral density of fluctuations,
$\langle \xi^2_\omega\rangle \equiv \int dt \langle\xi(t)\xi(0)\rangle e^{i\omega t}$.

For short-correlated noise this gives
\begin{equation}\label{eq:shortcorrnoise}
\langle \chi^2(t)\rangle \approx \langle\xi^2_{\omega=0}\rangle\cdot t \,,
\end{equation}
while for $1/f$ noise
$\langle\xi^2_\omega\rangle = \nu/|\omega|$ one finds
\begin{gather}
\langle \chi^2(t)\rangle \approx \frac{\nu}{\pi} t^2 \ln\frac1{\omega_\textrm{ir}t}
\,,
\end{gather}
where $\omega_\textrm{ir}$ is the infrared cutoff frequency.

Similar to the previous section, the overlap can be found from
\begin{equation}
P^2 = P_0^2 \sum_{q,q'} A^2_q A^2_{q'}
e^{-iv_\gr^0(q-q')\chi(t)} \,.
\end{equation}
For a Gaussian wave packet of width $w$, after summation over momenta and averaging over noise we find
\begin{equation}
P^2=P_0^2
\left(1+\frac{v_\gr^2}{w^2}
\langle \chi^2(t)\rangle \right)^{-1/2}
\label{eq:overlap_un_time}
\end{equation}
Note a similarity of Eqs.~\eqref{eq:overlap_un_time} and \eqref{eq:overlap}. In both cases fidelity is suppressed due to fluctuations of the velocity (or, equivalently, travel time). The case of short correlation time, Eq.~\eqref{eq:shortcorrnoise} is similar to $\delta$-correlations in space, Eq.~\eqref{eq:overlap} in terms of scaling with the travel time/length.

\begin{figure}
\includegraphics[width=0.8\columnwidth]{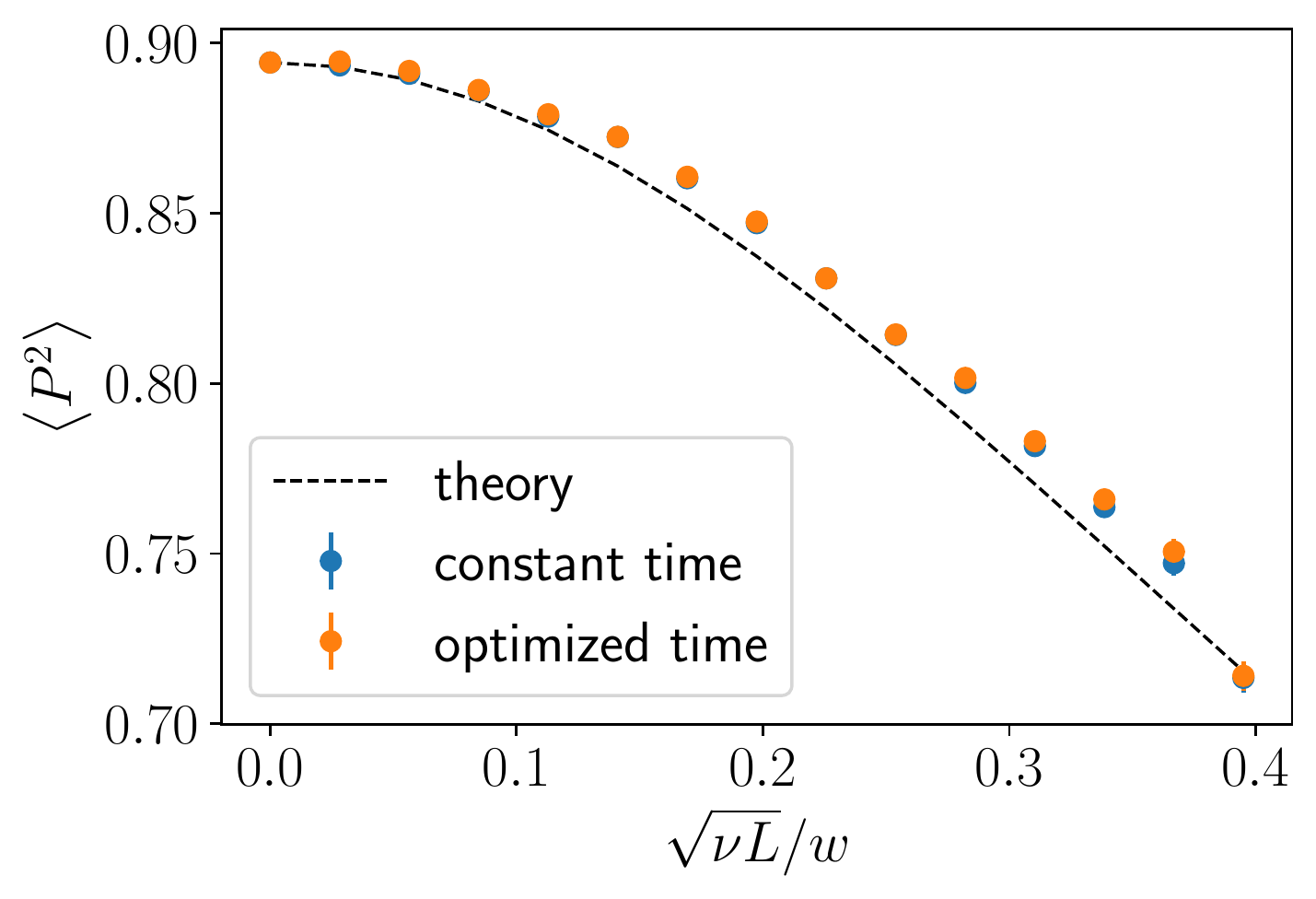}
\caption{\label{fig:Overlap_vs_delta_noise_an}
Overlap vs. disorder for $\delta$-correlated $1/f$ noise.
Notations and parameters are the same as in Fig.~\ref{fig:Overlap_vs_disorder_const} and $v_\gr/\omega_\textrm{ir}=19.1$. Dashed line follows Eq.~\eqref{eq:overlap_delta_1/f}.
}
\end{figure}

Consider now the effect of short-correlated low-frequency noise. We assume that relative fluctuations for each $ij$-coupling,
$\delta J_{ij}/J = \xi_{ij}(t)$ or $\delta \kappa_{ij}/\kappa = \xi_{ij}(t)$, are Gaussian, uncorrelated and have the same spectral density, $\langle (\xi_{ij})_\omega^2\rangle = \nu/|\omega|$.
Similarly to the Fig.~\ref{fig:Overlap_vs_disorder_const}  left, the  Fig.~\ref{fig:Overlap_vs_delta_noise_an} shows the numerically evaluated overlap vs. noise level for two detection methods: after the time span given by the nominal transport time in a clean system and after a time corrected for the noise-averaged wave packet delay ({\it optimized} time).

In this limit, the effect of the noise is similar to the stationary disorder in Section~\ref{sec:staticdisorder}, while in the evaluation of the local variance of the couplings the contribution of high frequencies should be neglected.
The dashed line in Fig.~\ref{fig:Overlap_vs_delta_noise_an} shows the expected value for the optimal-time approach:
\begin{equation}\label{eq:overlap_delta_1/f}
\langle P^2\rangle = P_{0}^{2}
\left( 1 + \frac{11}{9} \frac{L\nu}{w^2}  \ln\frac{v_\gr}{\omega_\textrm{ir}} \right)^{-1/2} \,.
\end{equation}

Note that for the numerical parameters corresponding to the Fig.~\ref{fig:Overlap_vs_delta_noise_an}, the average shift of the wave packet is much smaller than its fluctuations, hence the \emph{optimized} time is the same as the travel time for the clean system.

\section{Summary and discussion}
We discussed how chiral Majorana edge modes, specifically in Kitaev honeycomb model, can be used to transmit and process quantum information with the focus on the influence of various practical limitations and imperfections on fidelity of such operations. As an example, we considered an algorithm to perform the SWAP operation on two external qubits, coupled to the edge. This involves writing a quantum state onto the edge, propagation along the edge, and readout at a different location. To complement other approaches, we proposed an algorithm to transfer the quantum state of an external qubit to the chiral edge, which allows to create a wave packet of needed width with a fully controlled profile. Wider wave packets are more robust to distortion of their shape because of the nonlinearity of the edge mode spectrum.

We found that the ground state of the system remains vortex-free at not too strong disorder, demonstrating some stability. Furthermore, we verified that the edge modes demonstrate robustness towards disorder-induced localization. At the same time, their properties are affected by disorder or non-stationary noise, suppressing fidelity of quantum-state transmission along the edge. In a finite system this leads to fluctuations of the wave-packet propagation time, and thus limits reproducibility of the results between nominally identical samples.
We have found that homogeneous non-stationary fluctuations and $1/f$ spatially uncorrelated noise affect the fidelity in qualitatively similar ways due to locality of the propagating wave-packet.

Overall, the proposed algorithm allows to reach high fidelity 0.95, Eq.~\eqref{eq:fidelity} without further optimization of the SWAP operation, cf.~Fig.~\ref{fig:4_data}, and is stable with respect to considered imperfections of the lattice (Fig.~\ref{fig:Overlap_vs_disorder_const}).
The spread in the values of nominally identical circuit parameters on the order of 10\% should not prevent one from realization of the algorithm in a sample of several hundred qubits. This should be attainable, for instance, in superconducting-qubit networks~\cite{GoogleSupremacy} or other artifical systems.

From the practical point of view a relevant type of imperfection are vacancies in the lattice, which may host zero modes~\cite{KAO_vac, KAO2021168506}, and these will be analyzed in future work.

\section{Acknowledgments}
YM is grateful to A.~Wallraff for comments on disorder and noise levels in superconducting qubit networks.
This work has been supported by RFBR under No.~20-52-12034 and by the Basic research program of HSE.

\bibliographystyle{apsrev4-2}
\bibliography{references.bib} 
\end{document}